\documentclass[lettersize,journal]{IEEEtran}
\IEEEoverridecommandlockouts
\usepackage{makecell}
\usepackage{array}
\usepackage{graphicx,amssymb,amsmath}
\usepackage{multicol}
\usepackage[noadjust]{cite}
\usepackage{setspace}
\usepackage{subfigure}
\usepackage{graphicx}
\usepackage{float}
\usepackage{url}
\usepackage{stfloats}
\usepackage{amsthm,pifont}
\usepackage{flushend}
\usepackage{cases,subeqnarray}
\usepackage{bm,multirow,bigstrut}
\usepackage{amsmath, amsthm, amssymb}
\usepackage{textcomp}
\usepackage{latexsym,bm}
\usepackage{booktabs}
\usepackage{xcolor}
\usepackage{mathtools}
\usepackage{dsfont}
\usepackage{extarrows}
\usepackage{epsfig}
\usepackage{epstopdf}
\usepackage[noend]{algpseudocode}
\usepackage{algorithmicx,algorithm}

\usepackage{algorithm}
\usepackage{algpseudocode}
\usepackage{amsmath}
\usepackage{pifont}
\usepackage{cite}
\usepackage{bm}
\usepackage{cleveref}
\usepackage{multicol}       
\usepackage{multirow}       
\usepackage{array}          
\usepackage{colortbl}
\usepackage{makecell}
\definecolor{crimson}{RGB}{192,0,0}         
\definecolor{navy}{RGB}{47,85,151}         
\usepackage{bbding}
\usepackage{graphicx}
\usepackage{booktabs}
\usepackage{algorithm}
\usepackage{algpseudocode}

\theoremstyle{plain}

\theoremstyle{plain}

\usepackage{amsmath}

\IEEEoverridecommandlockouts
\begin{document}
\title{Stacked Intelligent Metasurfaces for 6G Wireless Networks: Principles, Applications, and Research Directions}

 \author{Enyu Shi, Jiayi Zhang,~\IEEEmembership{Senior Member,~IEEE}, Zhilong Liu, Ziheng Liu, Arumugam Nallanathan,~\IEEEmembership{Fellow,~IEEE}, M{\'e}rouane Debbah,~\IEEEmembership{Fellow,~IEEE}, Shi Jin,~\IEEEmembership{Fellow,~IEEE}, and Bo Ai,~\IEEEmembership{Fellow,~IEEE}

\thanks{E. Shi, J. Zhang, Z. Liu, Z. Liu, and B. Ai are with the State Key Laboratory of Advanced Rail Autonomous Operation, and also with the School of Electronics and Information Engineering, Beijing Jiaotong University, Beijing 100044, P. R. China. (email: jiayizhang@bjtu.edu.cn).}
\thanks{A. Nallanathan is with the School of Electronic Engineering and Computer Science, Queen Mary University of London, London E1 4NS, U.K. and also with the Department of Electronic Engineering, Kyung Hee University, Yongin-si, Gyeonggi-do 17104, Korea (email: a.nallanathan@qmul.ac.uk).}
\thanks{M. Debbah is with KU 6G Research Center, Department of Computer and Information Engineering, Khalifa University, Abu Dhabi 127788, UAE (email: merouane.debbah@ku.ac.ae) and also with CentraleSupelec, University Paris-Saclay, 91192 Gif-sur-Yvette, France.}
\thanks{S. Jin is with the National Mobile Communications Research Laboratory, Southeast University, Nanjing, 210096, P. R. China (email: jinshi@seu.edu.cn).}
}

\maketitle
\vspace{-1cm}
\begin{abstract}
The sixth-generation (6G) wireless networks are expected to deliver ubiquitous connectivity, resilient coverage, and intelligence-driven services in highly dynamic environments. To achieve these goals, distributed wireless architectures such as cell-free massive multiple-input multiple-output (MIMO) have attracted significant attention due to their scalability and fairness. Recently, stacked intelligent metasurfaces (SIMs) have emerged as a promising evolution of reconfigurable intelligent surfaces, offering multi-layer electromagnetic domain processing with enhanced controllability and spatial degrees of freedom. By integrating SIMs into distributed wireless networks, advanced wave-domain operations can be realized, enabling efficient interference management, improved energy and spectral efficiency, and robust physical-layer security. This article provides a comprehensive overview of SIM-aided distributed wireless networks, including their application scenarios, classification, and system architectures. Key signal processing challenges, such as hierarchical frameworks, user association, and joint precoding, are discussed, followed by case studies demonstrating significant performance gains. Finally, future research directions in hardware design, energy consumption modeling, algorithm development, and artificial intelligence integration are highlighted, aiming to pave the way for scalable and intelligent 6G distributed wireless networks.
\end{abstract}
\IEEEpeerreviewmaketitle


\section{Introduction}
The sixth-generation (6G) wireless networks are expected to enable ubiquitous connectivity, all-domain coverage, and intelligence-driven communication services for emerging applications such as autonomous transportation, industrial automation, and low-altitude networks \cite{ShiProc}. To meet these ambitious goals, future systems must achieve not only higher spectral efficiency (SE) and energy efficiency (EE) but also resilient and flexible coverage in dynamically changing environments. Traditional cellular architectures, however, suffer from limited cooperation and strong inter-cell interference. Consequently, distributed wireless architectures such as cell-free massive multiple-input multiple-output (CF mMIMO) and cooperative multi-point transmission save attracted growing attention as key enablers of 6G, offering enhanced scalability, improved user equipment (UE) fairness, and seamless mobility support through dense deployment of access points (APs) interconnected via high-capacity fronthaul networks.

To further enhance the capability of distributed wireless networks, reconfigurable intelligent surface (RIS) technology has emerged as a promising approach to reconfigure wireless propagation environments in a cost- and energy-efficient manner \cite{tang2020wireless}. By dynamically adjusting the phase of sub-wavelength elements, RIS can reshape incident electromagnetic (EM) waves to improve link quality and suppress interference. However, conventional single-layer RIS is limited to simple reflection operations with narrowband characteristics and restricted spatial degrees of freedom. As a result, much of the EM-domain processing potential remains unexploited in current 6G systems.

Motivated by these limitations, the concept of stacked intelligent metasurfaces (SIMs) has recently been introduced as a natural evolution of RIS technology \cite{liu2022programmable,an2024stacked}. A SIM consists of multiple programmable metasurface layers arranged in a cascaded manner with massive meta-atoms, enabling multi-layer wave manipulation through inter-layer diffraction and coupling. This volumetric architecture allows SIMs to perform advanced EM-domain operations such as multi-layer beam steering, spatial filtering, and near-field focusing \cite{papazafeiropoulos2024achievable}. Unlike single-layer RISs that merely reflect incident waves, SIMs act as programmable wave-domain processors capable of dynamically transforming amplitude, phase, and polarization. This unique capability fundamentally distinguishes SIMs from RISs, substantially enhancing channel controllability and information-processing capacity. When integrated with base stations (BSs), SIMs can offload part of the baseband digital precoding into the EM domain, simplifying hardware implementation and reducing computational burden \cite{an2023stacked}. Operating directly in the EM domain also enables near-light-speed signal processing and provides additional spatial degrees of freedom for efficient wave baseband co-design.

By leveraging these capabilities, integrating SIMs into 6G distributed wireless networks unlocks unprecedented opportunities for large-scale cooperative transmission and cross-domain optimization. Deployed near BSs/APs or within key propagation regions, SIMs can serve as cost-effective distributed nodes that reshape radio environments, mitigate multi-user interference, and extend coverage to low-altitude or obstructed areas. Furthermore, SIM-enhanced distributed architectures can reduce fronthaul overhead and enable hierarchical signal processing across the EM and digital domains, thereby improving both spectral and energy efficiency \cite{shi2025uplink}. With these unique advantages, SIMs are expected to become a cornerstone technology for future 6G distributed wireless networks, bridging the gap between EM intelligence and large-scale network cooperation.

This article provides a comprehensive overview of SIMs for 6G distributed wireless networks, covering their principles, applications, and research directions. Section~\ref{se:model} introduces representative application scenarios, SIM classifications, and system architectures. Section~\ref{se:challenges} presents the hierarchical signal processing frameworks under centralized, hybrid, and fully distributed configurations. Section~\ref{se:numerical} provides a case study to demonstrate the potential performance gains. Future directions regarding hardware design, energy modeling, low-complexity algorithms, and artificial intelligence (AI) integration are discussed in Section~\ref{se:future}, followed by conclusions in Section~\ref{se:conclusion}.

\begin{figure*}[!t]
\centering
\includegraphics[scale=0.65]{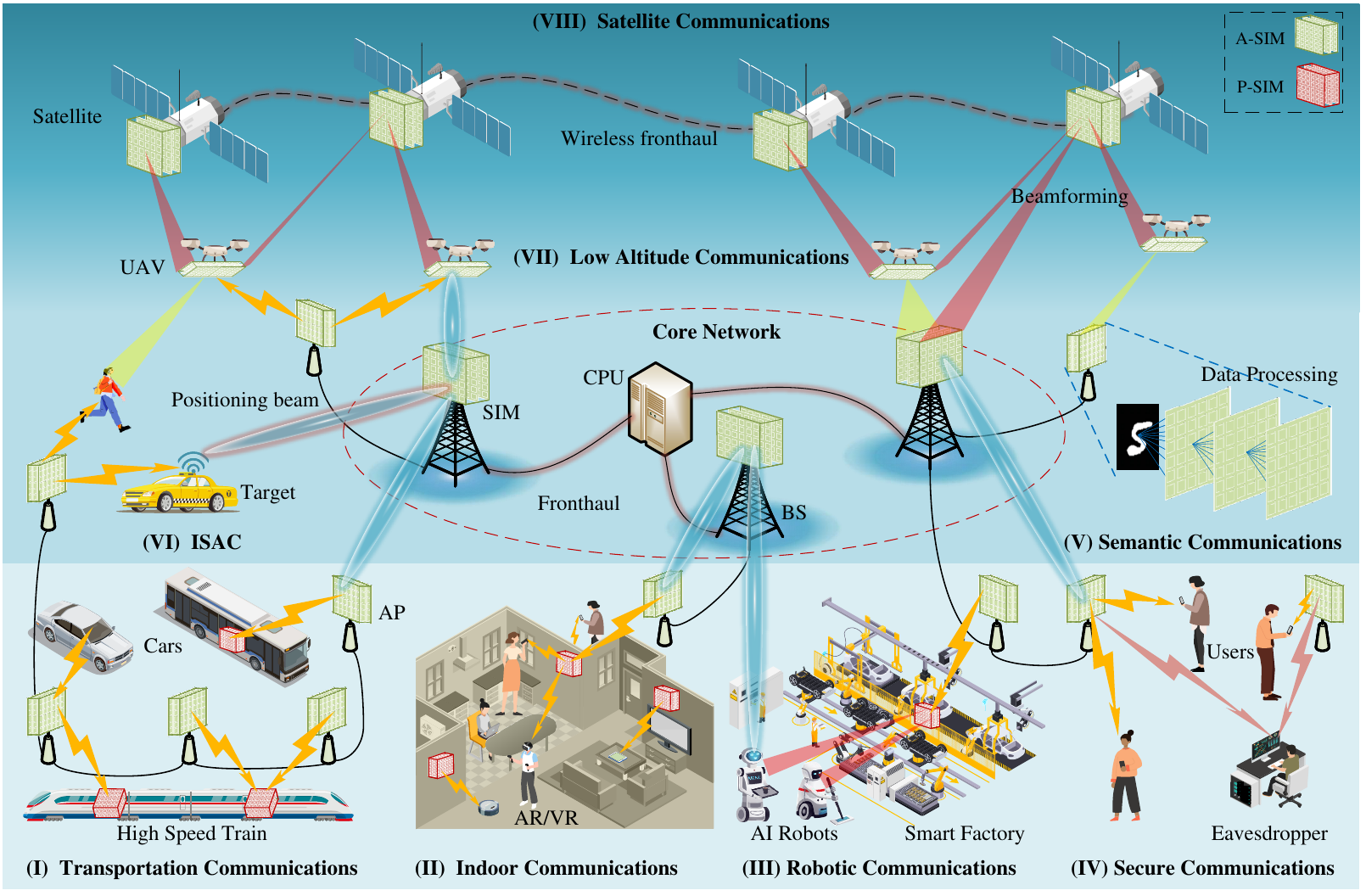}
\caption{Main application scenarios of SIM-aided 6G distributed wireless networks.}\label{Figure1}
\end{figure*}

\section{Main Application Scenarios, Classification, and System Architectures} \label{se:model}
In this section, we highlight the main application scenarios of SIM-aided distributed wireless networks. Then, we classify SIM based on system requirements and deployment characteristics. In addition, we elaborate on the representative architectures and illustrate the corresponding transmission procedures.

\subsection{Main Application Scenarios}\label{se:scenarios}
As shown in Fig.~\ref{Figure1}, SIMs are particularly suited to 6G distributed wireless networks where the propagation environment must be shaped in real time to sustain wide-area coverage, spatial selectivity, and low-latency control.
To better illustrate their deployment, SIMs can be classified into two categories: active SIMs (A-SIMs) and passive SIMs (P-SIMs). A-SIMs are tightly coupled with BS antennas, enabling real-time dynamic beamforming and effectively enhancing BS capabilities. In contrast, P-SIMs, similar to RISs, are usually deployed on walls, vehicles, or other structures. They consume less power while reshaping the EM environment and mitigating coverage holes.
In transportation communications, such as high-speed rail corridors and smart roads, channels suffer from rapid blockage and Doppler effects. Distributed SIMs deployed linearly along the track or roadside can cooperate with BS/APs to reshape beams, maintain link continuity, and suppress co-channel interference. Furthermore, P-SIMs mounted on vehicle bodies can reduce penetration loss, mitigating signal attenuation and improving the quality of experience for in-vehicle users.
Moreover, in indoor hotspot scenarios, such as those involving augmented reality (AR) and virtual reality (VR), SIMs can utilize near-field focusing and fine-grained spatial beamforming to scale capacity without proliferating RF chains.
In addition, for machine-dense scenarios such as smart factories and modernized farmlands, deploying distributed SIMs enables latency-critical control loops to leverage wave-domain processing that pre-shapes the EM environment around mobile robots and machinery, thereby stabilizing links under metallic clutter.
For secure communications, the SIM performs spatial filtering to reduce energy leakage toward potential eavesdroppers, thereby complementing baseband secrecy mechanisms \cite{10767193}. Moreover, the multilayer structure of the SIM can encrypt the signal layer by layer, such that at the receiver side, the authentic data can only be retrieved after the signal is sequentially decrypted through a matched SIM, thereby enhancing physical-layer security.
\begin{figure*}[!t]
\centering
\includegraphics[scale=0.9]{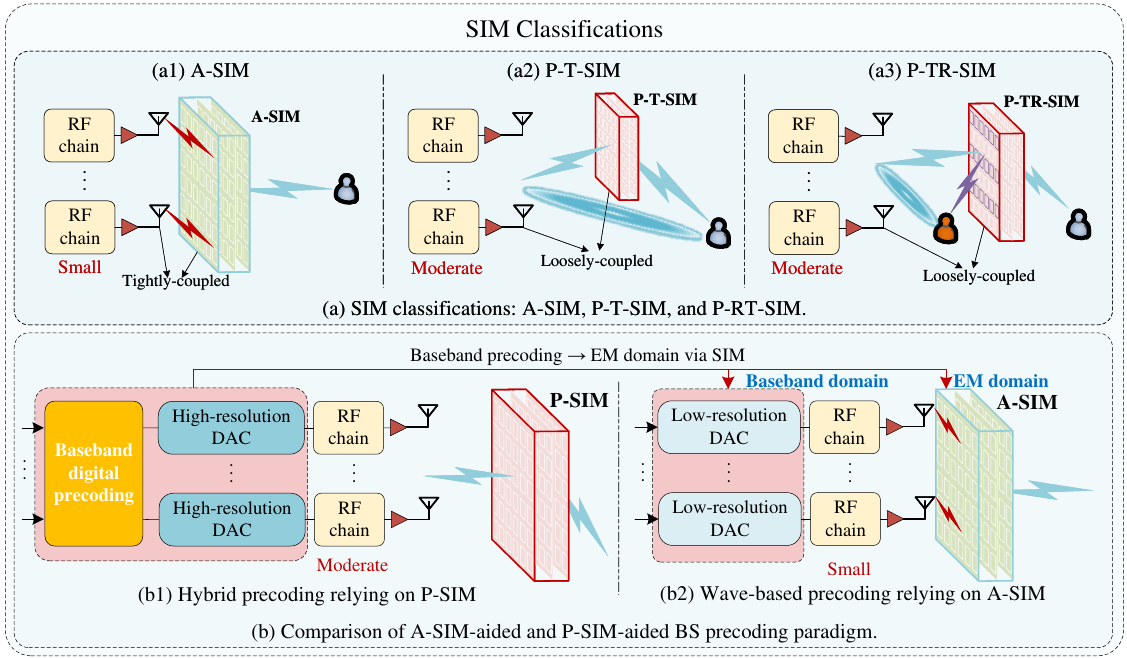}
\caption{SIM classifications and BS precoding paradigms. (a) Hardware taxonomy: A-SIM, P-T-SIM, and P-TR-SIM; (b) comparison of hybrid P-SIM aided precoding and A-SIM aided wave-domain precoding, highlighting the baseband EM split for A-SIM vs. P-SIM.} \label{SIM_Classification}
\end{figure*}
Beyond conventional communication, SIMs also support emerging paradigms. In semantic communication scenarios, SIMs enable direct extraction of semantic features in the EM domain, thereby eliminating redundant bit-level transmission and significantly reducing latency. Their multilayer structure enhances spectrum efficiency through effective information compression, while simultaneously enhancing robustness against interference and eavesdropping. Moreover, the scalable architecture of SIM supports low-complexity hardware implementation and can be readily extended to multimodal semantic tasks.
In integrated sensing and communication (ISAC) scenarios, distributed SIMs can jointly manipulate the propagation environment, improving both sensing accuracy and communication reliability \cite{10643881}.
For low-altitude scenarios such as UAV communications, SIM supports adaptive beamforming to maintain reliable links under high mobility, while its real-time wave-domain processing reduces RF chain power consumption, thereby extending UAV endurance.
Finally, integrating SIMs with satellite communications enables precise beam control, extending reliable services to obstructed or remote regions, including oceans and mountains, where terrestrial BSs are difficult to deploy.
Overall, these scenarios demonstrate that SIMs, when strategically deployed near BSs/APs or key propagation regions, can dynamically reshape the wireless environment in the wave domain, thereby complementing baseband cooperation to achieve reliable coverage, efficient interference management, and enhanced security.

\subsection{SIM Classification}\label{sec:sim_class}
\textbf{1) Hardware types and coupling:}
As shown in Fig.~\ref{SIM_Classification} (a), SIMs can be broadly classified into two functional categories: A-SIMs and P-SIMs. In this part, we provide a more detailed discussion of their respective functionalities and application scenarios.
\begin{itemize}
\item A-SIM: As shown in Fig.~\ref{SIM_Classification} (a1), an A-SIM is tightly coupled with the antenna array and performs wave-domain precoding in the EM domain. By migrating part of the baseband digital precoding functionality into the EM domain, each antenna only needs to transmit a single-user data stream, thereby simplifying the BS hardware. With moderate hardware complexity and power consumption, A-SIM achieves stronger beamforming capability and near light-speed computational efficiency through cross-domain design between the baseband and the EM domain. Typical deployment scenarios include BS front-ends requiring high beamforming precision.
\item P-T-SIM (Passive transmissive SIM): As shown in Fig.~\ref{SIM_Classification} (a2), a P-T-SIM is usually deployed at a certain distance from the BS, functioning as a passive transmissive panel similar to conventional RIS. By adjusting the phase of incident EM waves, it redistributes signal energy. Different from RIS, its multilayer structure enables more refined signal processing and precise beamforming, making it suitable for deployment on walls, windows, and vehicle bodies. It offers minimal hardware footprint, ultra-low power consumption, and flexible deployment for coverage enhancement and blind-spot mitigation, but the achievable gain is limited, and its noise resilience is relatively weak.
\item P-TR-SIM (Passive transmissive-reflective SIM):  As shown in Fig.~\ref{SIM_Classification} (a3), a P-TR-SIM is a passive structure that simultaneously supports both reflection and transmission of signals. In the first metasurface layer, some meta-atoms directly reflect incoming signals to serve nearby users, while the rest allow transmission. The transmitted signals then pass through subsequent stacked metasurfaces for further wave-domain beamforming, serving users on the opposite side. By moderately increasing complexity, P-TR-SIM achieves higher coverage flexibility, making it particularly effective in environments with users on both sides of the structure, such as corridors or canyon-like scenarios.
\end{itemize}

\begin{table*}[!t]
\centering
\caption{Different Signal Processing Frameworks for SIM-Aided Wireless Communication Networks}\label{Table1}
\renewcommand{\arraystretch}{1.3} 
\setlength{\tabcolsep}{6pt} 
\begin{tabular}{|c|c|c|c|c|c|}
\hline
\textbf{Signal Processing Framework} & \textbf{Ref.} & \textbf{Scenario} & \textbf{A/P-SIM} & \textbf{Objective} & \textbf{CSI} \\
\hline
\textbf{Centralized} &
\makecell{ \cite{an2023stacked} \\ \cite{10865993} \\ \cite{shi2025downlink} } &
\makecell{ HMIMO, DL \\ mMIMO, UL \\ CF mMIMO, DL} &
\makecell{ A-SIM \\ A/P-SIM \\ A-SIM } &
\makecell{ min channel error \\ max sum SE \\ max sum rate }  &
\makecell{ perfect \\ imperfect \\ perfect } \\
\hline
\textbf{Hybrid Centralized–Distributed} &
\makecell{ \cite{shi2025uplink} \\ \cite{li2024stacked} } &
\makecell{  CF mMIMO, UL \\ CF mMIMO, UL } &
\makecell{ A-SIM \\ A-SIM } &
\makecell{ max sum SE \\ max SE} &
\makecell{ imperfect\\ imperfect } \\
\hline
\textbf{Fully Distributed} & \cite{zhu2025joint} & CF mMIMO, DL & A-SIM & max sum SE  & imperfect\\
\hline
\end{tabular}
\end{table*}

\textbf{2) Relation to the precoding paradigm:}
Different types of SIMs impose distinct requirements on baseband and EM-domain signal processing, as illustrated in Fig.~\ref{SIM_Classification}(b). In hybrid precoding architectures with P-SIMs, since the panels are not tightly attached to the BS antenna array, there generally exists a direct path between the BS and the users that bypasses the SIM. As a result, most digital precoding functions must remain at the BS, while the P-SIMs provide only coarse wavefront control. This still requires digital precoding at the baseband, followed by transmission to the users, with the SIM mainly performing RIS-like phase adjustment. Consequently, stream-specific RF chains and high-resolution DACs are still necessary at the BS, although P-SIMs enable flexible deployment and extremely low power consumption.

By contrast, A-SIMs leverage stacked layers to migrate part of the digital precoding into the EM domain via wave-based processing. This reduces the BS reliance on high-resolution ADCs/DACs and allows each antenna to transmit only a single-user power stream. Ultimately, the joint design of baseband power control and EM-domain phase coordination not only expands the spatial degrees of freedom but also reduces BS front-end complexity for a given multiplexing target.

\subsection{System Architecture}
As shown in Fig.~\ref{Figure1}, SIM-aided distributed wireless networks follow a hierarchical architecture comprising the central processing unit (CPU), BSs, dense APs, dedicated SIM layers, and UEs. The core network provides global resource allocation and policy execution, while BSs not only deliver services but also coordinate a large number of lightweight APs. Specifically, APs handle user data transmission and reception, perform local signal processing, and connect to the core network via fronthaul links for data forwarding and information exchange. This distributed design enables wide-area, cell-free coverage. Between the RF front end and the physical environment, a SIM layer is introduced. These SIMs may either be colocated with BS/AP front ends or strategically deployed at remote positions, where they manipulate the amplitude and phase of EM waves to enable wave-based computation and beam control.

Compared with conventional distributed designs that consist only of the CPU-BS/AP-UE layers, the addition of the SIM layer fundamentally alters the system functionality and organization. It establishes an additional EM control domain alongside the digital baseband, enabling fine-grained wave manipulation near the point of propagation and providing new spatial degrees of freedom for coverage shaping and interference management. Meanwhile, the SIM layer introduces a lightweight control interface that supports fast configuration and basic synchronization with the AP layer, without enforcing a fixed partition of processing tasks. The key insight is that the SIM layer acts as an architectural component capable of both computation and control in the EM domain. Rather than replacing digital-domain functions, it enhances the hierarchical structure, ultimately forming a four-layer signal processing architecture of CPU-BS/AP-SIM-UE.

\begin{figure*}[!t]
\centering
\includegraphics[scale=0.8]{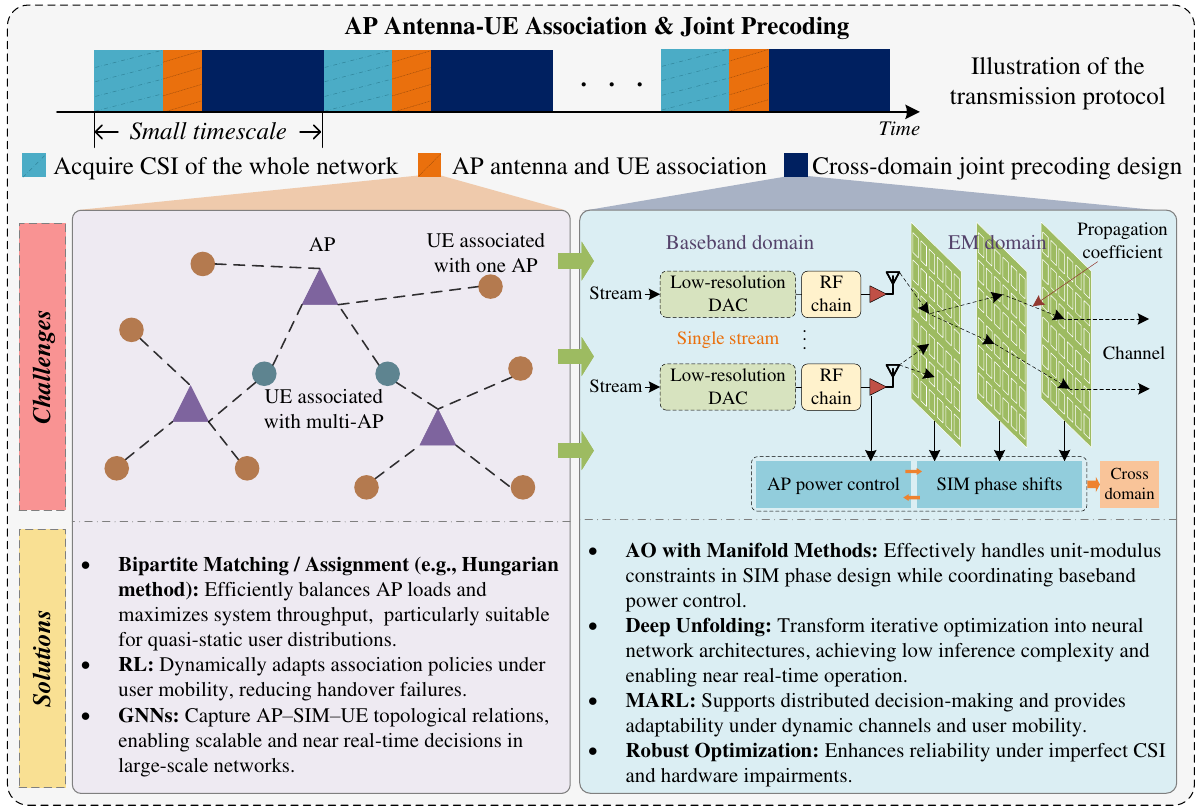}
\caption{Illustration of antenna–UE association and cross-domain joint precoding: challenges and potential solutions in SIM-aided distributed wireless networks.}\label{Association_precoding}
\end{figure*}

\section{Signal Processing Challenges}\label{se:challenges}
The integration of SIM into 6G distributed wireless networks introduces new signal processing challenges that must be addressed to fully exploit its potential. These challenges span from the design of scalable processing frameworks to user association and joint precoding optimization across both the baseband and EM domains.

\subsection{Signal Processing Framework}\label{se:framework}
Since the system follows a multilayer architecture, a key challenge is coordinating functionalities across components and performing hierarchical signal processing to achieve optimal system performance. Depending on the level of coordination, three representative signal processing frameworks can be identified as shown in Table \ref{Table1}.

\subsubsection{Centralized}
In centralized processing, the CPU has a global view of the network, collecting complete channel state information (CSI), allocating resources, and jointly optimizing the baseband and SIM configurations \cite{an2023stacked,10865993}. Although this approach maximizes performance, its reliance on high-capacity fronthaul links and centralized computation inevitably limits scalability and introduces significant latency \cite{shi2025downlink}.
\subsubsection{Hybrid Centralized-Distributed}
In the hybrid framework, distributed APs perform partial signal preprocessing and forward selected information to the CPU for global resource allocation, thereby balancing performance and overhead \cite{shi2025uplink,li2024stacked}. Compared with fully centralized processing, this approach reduces signaling overhead and computational burden while still benefiting from network-wide coordination. Two key open issues in this framework are: whether SIM control should be centrally managed by the CPU or delegated to local APs, and how much information each AP should forward to the CPU.
\subsubsection{Fully Distributed}
In fully distributed processing, each AP makes precoding and control decisions solely based on local CSI, and the associated SIM is controlled locally by the AP. This approach minimizes fronthaul load, significantly improves scalability, and reduces computational burden and latency. However, its limited coordination may lead to suboptimal interference management.

A promising solution is to design adaptive hierarchical frameworks that dynamically adjust the degree of centralization based on network conditions. For example, when fronthaul capacity and computational resources are sufficient, the system can operate in a more centralized mode to exploit global CSI. Conversely, under stringent latency or fronthaul constraints, it can shift toward a distributed mode, where APs and SIMs make local decisions with lightweight coordination. To further enhance hybrid operation, consensus-based mechanisms or the alternating direction method of multipliers (ADMM) can serve as system-level coordination tools, enabling APs and SIMs to iteratively exchange limited information and approximate centralized performance at lower overhead. In addition, embedding advanced learning-based methods such as federated learning for distributed cooperation or multi-agent reinforcement learning (MARL) for multi-agent coordination-can further improve scalability and adaptability, making the framework robust across diverse deployment scenarios \cite{zhu2025joint}.

\subsection{Antenna-UE Association}
As shown in Fig.~\ref{Association_precoding}, UE association is a fundamental prerequisite for reliable data transmission. SIMs are primarily introduced to offload baseband processing at the BS. By enabling near light-speed wave-domain signal processing, they relax the requirements on digital precoding and allow each antenna to transmit only a single-user data stream. However, this design raises a critical challenge: the antenna-UE association problem, as the number of AP antennas is often smaller than the number of UEs. In distributed networks, overlapping AP coverage further complicates this issue, requiring careful allocation of service UEs to fully utilize network resources and achieve optimal performance. Moreover, the inherent mobility of UEs requires that SIM-aided systems dynamically assign antennas and SIM resources under interference and load-balancing constraints. Hence, efficient association strategies are indispensable for ensuring fairness, maximizing spatial multiplexing gains, and fully leveraging SIM's wave-domain processing capabilities, which ultimately determine both user quality of service and network scalability.

A feasible solution is to formulate antenna-UE association as a bipartite matching or assignment problem, where algorithms such as the Hungarian method or maximum-weight matching can efficiently balance AP loads while maximizing overall throughput. In dynamic scenarios with high user mobility, learning-based approaches such as reinforcement learning can adapt association policies over time, leveraging mobility patterns to reduce handover failures. Moreover, graph neural networks (GNNs) can naturally capture the topological relationships among APs, SIMs, and UEs, enabling scalable and near-real-time association decisions.

\subsection{Joint Precoding Design}
The stability and efficiency of high-capacity data transmission in SIM-aided distributed networks critically depend on the design of joint precoding across the baseband and EM domains. In contrast to conventional architectures, where all digital precoding is executed at the BS, the use of SIMs introduces an additional wave-domain dimension, enabling part of the precoding process to be migrated into the EM layer as shown in Fig~\ref{Association_precoding}. This cross-domain feature significantly increases spatial degrees of freedom and opens new opportunities for interference suppression and coverage shaping \cite{an2023stacked}. However, it also brings fundamental challenges, as the coordination among power control, digital baseband precoding, and SIM phase configuration becomes highly coupled and nontrivial.

For P-SIMs, the design primarily focuses on phase-shift optimization in conjunction with conventional AP digital precoding. However, in distributed networks, P-SIMs are often not tightly bound to a specific AP, raising the question of which AP should control their beamforming and how this should be coordinated with BS operations. This becomes particularly challenging when fine-grained control must be ensured under limited feedback and imperfect CSI.
By contrast, A-SIMs leverage multilayer structures to enable more advanced wave-based precoding, thereby reducing the dependence on high-resolution DAC/ADCs and RF chains at the BS. Yet, this advantage comes at the cost of higher hardware complexity, stringent dynamic calibration, and increased synchronization overhead with the BS. Moreover, in distributed networks, how multiple AP-associated A-SIMs can cooperate to deliver coordinated multi-point services remains an important open challenge.

Developing low-complexity yet near-optimal joint precoding algorithms remains a central challenge in SIM-aided distributed networks. The difficulty is further compounded by practical constraints such as partial or outdated CSI, user mobility, hardware impairments, and fronthaul limitations.
To address these issues, algorithmic approaches have been proposed as shown in Fig~\ref{Association_precoding}. Alternating optimization (AO) combined with manifold-based methods can efficiently deal with the nonconvex unit-modulus constraints in SIM phase design while coordinating with baseband power allocation. Deep unfolding frameworks can approximate iterative optimization with low inference complexity, enabling real-time operation.
MARL supports distributed decision-making and provides adaptability under dynamic channels and user mobility. In addition, robust optimization can mitigate the effects of imperfect CSI and hardware impairments. These model- and learning-based methods complement each other, forming a pathway toward practical cross-domain precoding solutions that balance performance and complexity.

\begin{figure}[!t]
\centering
\includegraphics[scale=0.55]{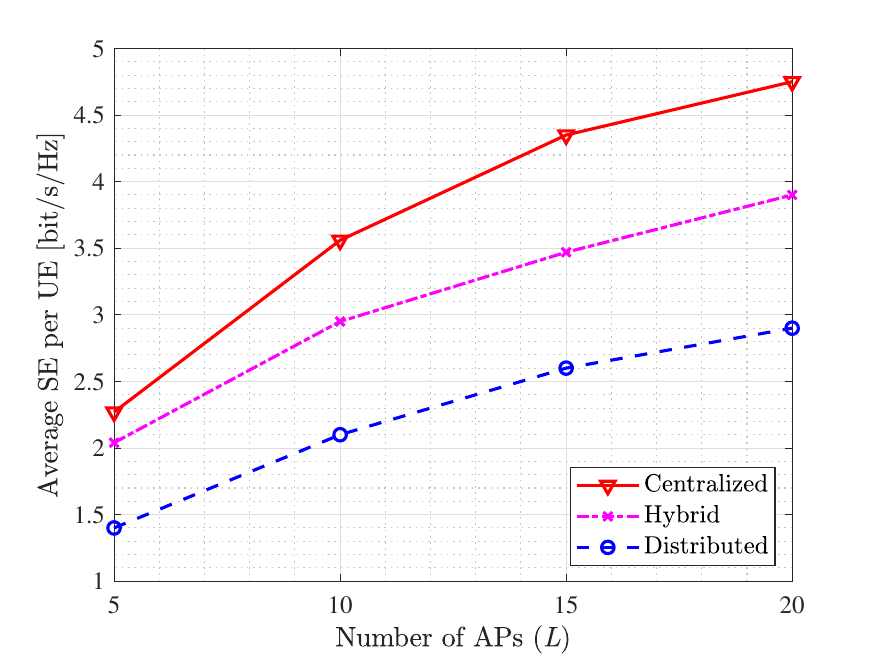}
\caption{Average SE per UE versus the number of APs for the uplink under different signal processing frameworks (2 antennas per AP, 5 layers per SIM, 64 meta-atoms per layer, and 5 UEs).} \label{Signal_processing_framework}
\end{figure}

\section{Case Study}\label{se:numerical}
In this section, we present case studies that demonstrate the benefits of deploying SIM in distributed wireless networks. Specifically, we consider a SIM-aided CF mMIMO network and analyze the system performance under different signal processing frameworks and various joint precoding schemes \cite{shi2025uplink}. The small-scale fading is assumed to be Rayleigh fading.
First, Fig.~\ref{Signal_processing_framework} illustrates the average SE versus the number of APs for the uplink under different signal processing frameworks. Centralized refers to the scheme where global CSI is forwarded to the CPU for centralized processing and decoding. The hybrid scheme means that signals are first locally preprocessed at the APs and then transmitted to the CPU for final decoding. In contrast, the distributed scheme indicates that each AP independently decodes the received signals, similar to a cell-based architecture. The results show that the centralized framework achieves the best performance, followed by the hybrid framework, while the fully distributed framework yields the lowest performance. As discussed earlier, centralized algorithms consume more transmission resources and computational power to deliver higher performance, whereas fully distributed algorithms suffer from performance degradation because decisions are made locally. The hybrid framework for SIM-phase design and AP power control achieves a balance by enabling APs to perform local signal preprocessing before forwarding the results to the CPU for centralized decoding, thereby effectively trading off complexity and performance. Therefore, in practical systems, different signal processing frameworks should be flexibly adopted depending on the specific requirements.

\begin{figure}[!t]
\centering
\includegraphics[scale=0.55]{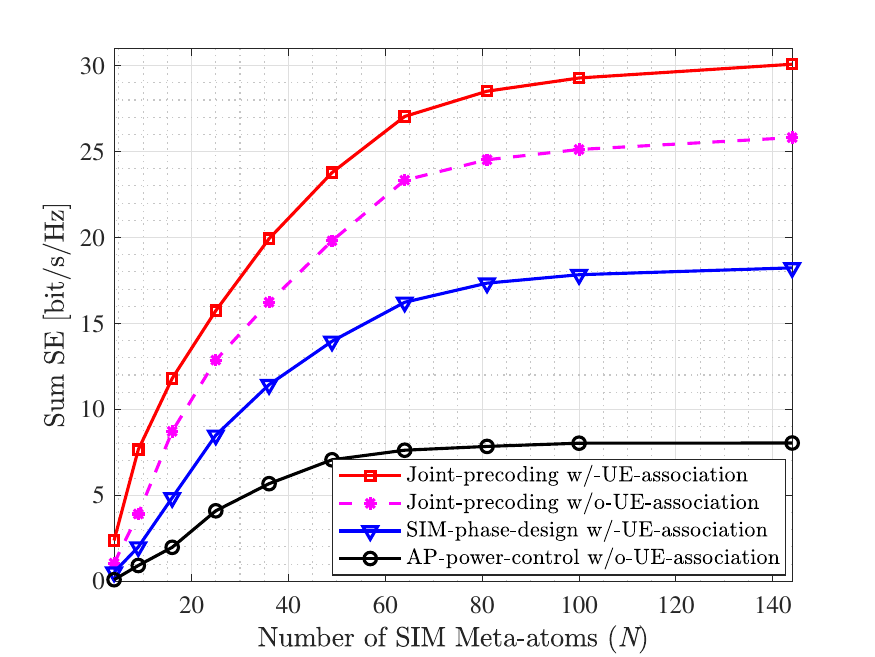}
\caption{Sum SE performance under different precoding schemes for the downlink of the CF mMIMO network (4 APs, 2 antennas per AP, and 2 layers per SIM).}\label{Joint_precoding}
\end{figure}

Fig.~\ref{Joint_precoding} shows the sum SE under different precoding schemes for the downlink of the CF mMIMO network. Joint-precoding w/-UE-association means jointly designing the AP transmit power and SIM phase shifts while taking antenna-UE association into account. SIM-phase-design means only optimizing the SIM phase shifts with equal power allocation, and AP-power-control means only designing the AP transmission power with fixed SIM phases. The results show that performance improves as the number of meta-atoms increases; however, when the number of meta-atoms becomes very large, further increases yield diminishing returns. Once the performance stabilizes, the joint-precoding scheme outperforms the SIM-phase design and AP-power-control benchmarks by 67\% and 275\%, respectively. Moreover, the UE-association design achieves a 16\% performance improvement compared with random UE access. These results validate the critical role of UE association and joint precoding in enhancing the performance of SIM-aided distributed wireless networks.

\section{Future Directions}\label{se:future}
SIM-aided distributed wireless networks offer broad prospects for future development. In the following, we list several promising research directions.
\subsection{Hardware Design}
The practical deployment of SIMs requires breakthroughs in hardware design, particularly in metasurface fabrication, integration with RF front ends, and scalable control circuits. Current prototypes are still constrained by limited reconfigurability, slow tuning speed, and low power efficiency (e.g., significant transmission loss), which restrict their applicability in large-scale distributed networks. An important research direction is leveraging advanced materials to balance complexity and performance, thereby enabling the development of low-cost, energy-efficient, and high-precision SIM units capable of wideband operation under high-mobility conditions.

Furthermore, integration with BSs/APs introduces new demands for compact packaging, thermal stability, and robust synchronization. Future designs should investigate hybrid active passive structures, exploit advanced materials such as tunable semiconductors, and adopt flexible architectures that support multi-band and multi-functional operation. These innovations will be essential to bridge the gap between theoretical concepts and practical large-scale SIM deployments.

\subsection{Energy Consumption Modeling}
EE is a key concern for SIM-aided networks, especially as dense deployments and multi-layer SIMs may significantly increase overall power consumption. Existing models often oversimplify by ignoring hardware power consumption, such as the power consumption of the SIM controller and the dynamic switching energy of meta atoms, leading to inaccurate system-level evaluations \cite{EESIM}. Therefore, accurate and tractable energy consumption models that capture both static and dynamic components are urgently needed.

In distributed architectures, where large numbers of SIMs and APs cooperate, energy-aware frameworks should jointly consider SE, coverage, and energy cost across multiple layers. Future work should explore cross-layer energy models, joint power control with SIM reconfiguration, and sustainable designs leveraging renewable energy or wireless power transfer. Such approaches are critical to enabling green, scalable, and cooperative SIM-aided distributed networks.

\subsection{Low-Complexity Algorithms}
The introduction of SIM adds a new EM dimension to the signal processing pipeline, greatly expanding the spatial degrees of freedom but also increasing computational complexity. Joint optimization of baseband precoding, SIM configuration, and power allocation is often formulated as a large-scale, highly nonconvex problem, which is computationally prohibitive in real-time systems.

Future work should aim at exploiting problem structure, hierarchical optimization, and approximation techniques to reduce complexity. In particular, distributed and hybrid schemes that decompose global optimization into smaller local tasks, while maintaining limited coordination among APs and SIMs, are especially promising for scalability. Moreover, the tradeoff between algorithmic complexity and system performance must be carefully quantified to guide practical design choices for large-scale distributed networks.

\subsection{Integration with Large AI Models}
The recent emergence of large AI models (LAMs) presents new opportunities for optimizing SIM-aided distributed wireless networks \cite{zhang2025multi}. Unlike conventional task-specific learning methods, LAMs trained on diverse datasets can generalize across heterogeneous environments, enabling more robust decision-making for user association, channel estimation, and joint precoding. By leveraging their reasoning and transfer capabilities, LAMs can act as unified controllers to coordinate multiple APs and SIMs, thereby enhancing adaptability in dynamic distributed environments.

However, integrating LAMs into distributed systems also introduces challenges, including latency, inference cost, and energy overhead. Large-scale deployment further requires lightweight model adaptation, privacy-preserving training, and efficient knowledge transfer across heterogeneous devices. Future research should explore model compression, federated fine-tuning, and hybrid approaches that combine domain-specific optimization with the generalization ability of LAMs. Such integration holds great potential to enable autonomous, scalable, and intelligent SIM-aided distributed wireless networks.

\section{Conclusions}\label{se:conclusion}
This article provided a comprehensive overview of SIMs for 6G wireless networks. By exploiting their multi-layer EM-domain processing capability, SIMs extend beyond conventional RIS, enabling advanced wave-baseband co-design that enhances SE and EE, improves coverage, suppresses interference, and strengthens physical-layer security. We discussed representative application scenarios, SIM classifications, system architectures, and key signal processing challenges, including hierarchical frameworks, antenna-UE association, and joint precoding. Case studies further demonstrated the significant performance gains achievable with SIM integration. Looking forward, future research should prioritize practical hardware design, accurate energy consumption modeling, low-complexity algorithms, and LAM-driven optimization to enable scalable, intelligent, and sustainable SIM-aided wireless networks for the 6G era.

\bibliographystyle{IEEEtran}
\bibliography{IEEEabrv,Ref}






\end{document}